%% file: cygnus_legacy_survey.tex
\documentclass{emulateapj}
\pdfoutput=1

\newcommand{\arcs}{\mbox{\ensuremath{^{\prime\prime}}}}
\newcommand{\arcm}{\mbox{\ensuremath{^{\prime}}}}

\usepackage{natbib}
\usepackage{psfig}
\usepackage{graphicx}

\shorttitle{{\it Chandra} Cygnus~OB2 Legacy Survey Catalog}
\shortauthors{Wright et al.}

\begin{document} 

\title{The {\it Chandra} Cygnus~OB2 Legacy Survey: Design and X-ray Point Source Catalog}

\author{Nicholas J. Wright$^{1,2}$, Jeremy J. Drake$^1$, Mario G. Guarcello$^1$, Tom L. Aldcroft$^1$, Vinay L. Kashyap$^1$, Francesco Damiani$^3$, Joe DePasquale$^1$, and Antonella Fruscione$^1$}
\affil{$^1$Center for Astrophysics, 60 Garden Street, Cambridge, MA~02138, USA}
\email{nick.nwright@gmail.com}
\affil{$^2$Centre for Astrophysics Research, University of Hertfordshire, Hatfield AL10 9AB, UK}
\affil{$^3$INAF - Osservatorio Astronomico di Palermo, Piazza del Parlamento 1, I-90134 Palermo, Italy}


\begin{abstract}

The Cygnus~OB2 association is the largest concentration of young and massive stars within 2~kpc of the Sun, including an estimated $\sim$65 O-type stars and hundreds of OB stars. The {\it Chandra} Cygnus~OB2 Legacy Survey is a large imaging program undertaken with the Advanced CCD Imaging Spectrometer onboard the {\it Chandra X-ray Observatory}. The survey has imaged the central 0.5~deg$^2$ of the Cyg~OB2 association with an effective exposure of $\sim$120~ks and an outer 0.35~deg$^2$ area with an exposure of $\sim$60~ks. Here we describe the survey design and observations, the data reduction and source detection, and present a catalog of $\sim$8,000 X-ray point sources. The survey design employs a grid of 36 heavily ($\sim$50\%) overlapping pointings, a method that overcomes {\it Chandra}'s low off-axis sensitivity and produces a highly uniform exposure over the inner 0.5~deg$^2$. The full X-ray catalog is described here and is made available online.

\end{abstract}


\section{Introduction}

The {\it Chandra X-ray Observatory} \citep{weis02} is the premier X-ray observatory for studying young star clusters and associations, thanks to its low background rate and high angular resolution, and the ideal choice for a wide survey of the Cygnus~OB2 association \citep[e.g.,][]{mass91,knod00,wrig10a}. The {\it Chandra} Cygnus~OB2 Legacy Survey \citep{drak14} is a 1~deg$^2$ survey of the Cyg~OB2 association with the {\it Chandra X-ray Observatory} that goes deeper and wider than previous surveys \citep{alba07,wrig09a}. \citet{drak14} discuss the survey motivation and design and present highlights of the results. This paper describes the survey observations, X-ray data reduction, and the compilation of an X-ray point source catalog. An analysis of the completeness limits of the survey as a function of various observational and stellar parameters is presented in \citet{wrig14e}. Other papers in this series present a catalog of optical and near-IR counterparts to the X-ray catalog \citep{guar14} and an analysis of Cyg~OB2 members and contaminants in the X-ray catalog \citep{kash14}.

In Section~2 we introduce the survey, describe its design and present the observations. In Section~3 we present the methods used to detect sources in the survey, including the use of a new enhanced wavdetect code. In Section~4 we outline the iterative process used to refine the list of reliable sources and the methods used to extract X-ray source properties from the data. Finally in Section~5 we present an analysis of the quality of the final X-ray source catalog. Future papers will include X-ray spectral fitting of the brightest sources detected \citep{flac14b} and will combine this X-ray catalog with the available optical and infrared data covering the region to produce a fully multi-wavelength catalog of Cyg~OB2 members.

\section{Observations}

In this section we describe the survey strategy, observations, and initial data reduction techniques employed. The survey was selected in 2009 as a {\it Chandra X-ray Observatory} Cycle~11 Very Large Project (VLP) and awarded 1.08~Ms. The observations presented here include the 36 pointings that make up the survey, as well as 4 previous observations of Cyg~OB2. These observations have been published elsewhere, but we include them in our data reduction and analysis so as to provide a single cohesive dataset. We will assume that all X-ray sources in Cyg~OB2 are at a distance of 1.4~kpc, as determined by recent parallax measurements of masers within the Cygnus~X complex \citep{rygl12}.

\subsection{Survey Design}
\label{s-design}

The {\it Chandra} Cygnus~OB2 Legacy Survey was designed to uniformly survey the entire OB association to a high depth that would be sufficient to identify  a large population of young, solar mass stars that would facilitate unbiased surveys of the structure \citep[e.g.,][]{wrig14b} and stellar populations \citep{guar13} of the association. Given the large size of the Cyg~OB2 association \citep[][estimate a half-light radius of 13\arcm\ and a total diameter of $\sim$2$^\circ$]{knod00} this necessrily required multiple pointings. To minimize the effects of {\it Chandra}'s reduced point spread function (PSF) sensitivity at large off-axis angles a simple tiling strategy was adopted \citep[following that used successfully by][]{elvi09}, that has been shown to produce a well-defined lower flux limit with a sharp cutoff \citep{pucc09} and a high spatial resolution over the majority of the survey area. This approach also ensures that the area with {\it Chandra}'s good PSF, which can resolve sources 2\arcs\ apart (corresponding to $\sim$0.01~pc at 1.4~kpc) is maximized. This has allowed us to obtain a spatially-unbiased and accurate census of the association facilitating a number of scientific studies of interest. The sensitivity of the survey resulting from our tiling strategy is discussed and simulated in detail in \citet{wrig14e}.

\begin{figure*}[t]
\includegraphics[width=520pt]{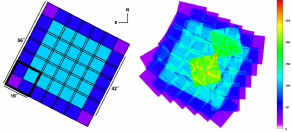}
\caption{{\it Left:} the ``as designed'' Cygnus~OB2 {\it Chandra} Legacy Survey tiling for the 36 30~ks pointings. The thick black box (bottom left) represents one ACIS-I pointing, the thin boxes all the pointings. Different colors show areas with different numbers of overlapping pointings: teal -- 4 overlapping pointings; blue -- 2 overlapping pointings; purple -- 1 pointing. The black bars show roughly the relative dimensions of one pointing ($\sim$16\arcm), of the deep inner area ($\sim$42\arcm), and of the total field ($\sim$56\arcm). Pointing 1-1 lies at the bottom right (SW), 1-6 lies at the top right (NW), 6-1 lies at the bottom left (SE) and 6-6 lies at the top left (NE). {\it Right:} the ``as executed'' exposure map for the Cygnus~OB2 {\it Chandra} Legacy Survey and complementary observations. The color bar gives the achieved effect exposure in units of ks. The deepest region of the survey with a non-negligible area has an exposure time of $\sim$220~ks.}
\label{exposuremap}
\end{figure*}

The tiling scheme (Figure~\ref{exposuremap}) employs a $6 \times 6$ raster array of 36 pointings using {\it Chandra}'s Advanced CCD Imaging Spectrometer \citep[ACIS-I;][]{garm03}, each of 30~ks nominal exposure. The center of the array, 20$^h$~33$^m$~12$^s$ +41$^\circ$~19\arcm~00\arcs, was chosen based on the main concentration of OB stars found by multiple authors. The 8\arcm.0 offset between pointing centers was chosen to be slightly less than the 8\arcm.3 size of an ACIS chip so that chip gaps are not co-added to create small scale dips in the effective exposure time. The inner part of the field is covered by four exposures, to give a total nominal exposure of $\sim$120~ks over a $42\arcm \times 42\arcm$ area (0.5~deg$^2$, hereafter referred to as the {\it deep inner} region). The outer regions are covered by two observations, and the four corners covered by one observation. The final position and overall position angle of the array was chosen to maximize the extent of the association covered (as traced by the positions of the known OB stars) as well as maximize the alignment of the different observations.

\subsection{Survey and Supplemental Observations}

The survey observations were performed over a 6-week period from January -- March 2010. All observations utilize the ACIS in imaging mode, which comprises four CCDs (chips I0-I3), each with $1024 \times 1024$ pixels (at a scale of 0.492\arcs\ pix$^{-1}$), giving a $17\arcm \times 17\arcm$ field of view (FoV). Some of the ACIS-S chips were turned on during the observations, but due to the high off-axis angle to these chips and consequently large PSF we have not used this data in this work. All observations were performed in {\sc very faint} mode. The indices $X-Y$ (1-1 through 6-6) describe the field numbers, where $X$ is an index in R.A. and $Y$ an index in declination, and 1-1 being in the bottom right (SW) corner of the grid and 1-6 being in the top right (NW) corner. The pointings and overall survey grid were observed at a nominal roll angle of 27.2$^\circ$, with the majority of pointings within the central grid of $4 \times 4$ pointings stable to $\pm 10$~deg (see Figure~\ref{exposuremap}).

None of the 30~ks observations were scheduled to be split due to observational or thermal constraints on the spacecraft, but the first observation, ObsID~10939, was interrupted after 24~ks for a spacecraft software reboot and the remaining 6~ks were observed as ObsID~12099. The mean effective exposure time per pointing (not per ObsID) for the 36 survey pointings is 28.1~ks, when only the good time intervals (GTIs) are used. The minimum and maximum exposure times per pointing are 27.3~ks and 30.6~ks. Over the inner region of $42\arcm \times 42\arcm$ the mean total exposure time per tile is 116.3~ks, with a variation of $\pm 0.7$~ks (0.6\%).

\input{obsid_table.tex}

These observations were combined with four existing observations that fell within the survey area and which had previously been used to study the Cyg~OB2 association and some of its members by other authors \citep{butt06,alba07,wrig09a}. In the center of the association this results in a maximum exposure time of a non-negligible area of $\sim$214~ks. In total 41 ObsIDs were used for this work, and these are listed in Table~\ref{obsids}.

\subsection{Data Processing}
\label{s-dataprocessing}

The data from all 41 ObsIDs were uniformly processed using the CIAO~4.5 software tools\footnote{http://cxc.harvard.edu/ciao/} \citep{frus06}, the {\it yaxx}\footnote{http://cxc.harvard.edu/contrib/yaxx/} tool, the {\it pyyaks}\footnote{http://cxc.harvard.edu/contrib/pyyaks/} tool, and the CALDB~4.5.8 calibration files. The standard Level-1 and Level-2 data products were downloaded from the {\it Chandra} Data Archive for all ObsIDs. After the data were processed we determined astrometric corrections (Section~\ref{s-astrometric}) and then reprocessed the data as outlined here.

Data reduction began with the Level~1 event files using the CIAO {\sc acis\_process\_events} tool to perform background cleansing and gain adjustments. A new Level~2 event file was produced by filtering out events with non-zero status and bad grades (events with grades 1, 5, or 7 were removed). While running the {\sc acis\_process\_events} tool we enabled very-faint mode processing and turned off pixel randomisation, applying instead the sub-pixel EDSER (Energy-Dependent Subpixel Event Repositioning) algorithm, which should result in the optimal event positions.

Intervals of high background were determined by creating a background light curve for the ACIS-I events with point sources found by {\sc wavdetect} removed. No observations showed intervals with a significant deviation from the quiescent background level (with the exception of those that included the bright X-ray source Cyg~X-3, see below). The background is very stable for the observations that avoid Cyg~X-3 with a typical rate of $\sim 4.9 \times 10^{-7}$ counts s$^{-1}$ pixel$^{-1}$. In the 5 pointings (6 ObsIDs: 10939, 10940, 10964, 10969, 10970, 12099) that include Cyg~X-3 the background is higher and has significant spatial variability due to the wings of the PSF from Cyg~X-3. Background rates in these ObsIDs are typically (1--$4) \times 10^{-6}$ counts s$^{-1}$ pixel$^{-1}$, which significantly reduces the sensitivity in these areas.

\subsection{Astrometric Corrections}
\label{s-astrometric}

The absolute astrometry provided by the {\it Chandra} spacecraft is accurate to one ACIS-I pixel (0.6\arcs at 90\% confidence, POG\footnote{{\it Chandra} Proposers Observatory Guide: http://cxc.cfa.harvard.edu/proposer/POG/}, Section~5). To avoid a loss of sensitivity when merging data from different observations, and to provide the most accurate positions for cross-matching to other wavelengths, the astrometry must be corrected to a common frame of reference. To do this a list of bright X-ray sources for each of the 41 ObsIDs was generated using the standard CIAO {\sc wavdetect} tool, considering only bright sources ($\geq 10$ photons) on-axis (within 4\arcm\ of the aim point). This list was then cross-matched with the 2MASS \citep[Two Micron All Sky Survey,][]{skru06} point source catalog using a matching radius of 1\arcs\ and using only sources with `AAA' quality photometry and errors $< 0.1$~mag in all three bands. Only sources in the magnitude range $K_s = 8$--14~mag were used, which minimizes systematic effects introduced by bright stars (saturation) and faint background objects (misidentification).

From the list of cross-matched sources, which typically contained between $\sim$20 and $\sim$100 sources per ObsID, astrometric offsets were calculated in RA and Dec. The spacecraft roll angle is known to a very high precision based on guide stars spaced a degree apart or more on the sky, and consequently astrometric uncertainties arising from roll angle uncertainties are negligible compared to those arising from knowledge of absolute pointing. Roll angle changes were therefore set to zero for all ObsIDs for the purposes of astrometric corrections. The offsets between the X-ray and near-IR positions have mean values of $\Delta$R.A.~$= 0.04$\arcs\ and $\Delta$Dec.~$= 0.03$\arcs\, with all values smaller than $\sim$0.1\arcs. The astrometric offsets were applied to the L1 data products to create new aspect solutions for each ObsID and the data were then re-reduced following the procedure in Section~\ref{s-dataprocessing}.

\subsection{Exposure maps and survey sensitivity}
\label{s-exposure}

Exposure maps were constructed for each ObsID on a per-CCD basis and in multiple energy bands using the standard CIAO tool sequence of {\sc asphist}, {\sc mkinstmap}, and {\sc mkexpmap}. The exposure maps were calculated using a thermal plasma model spectrum with $kT = 1.35$~keV and $N_H = 1.25 \times 10^{22}$~cm$^{-2}$ (typical of a stellar coronal X-ray source at the extinction of Cyg~OB2). Figure~\ref{exposuremap} shows the exposure map, which clearly shows the central region composed of four overlapping pointings and complemented by the two existing deep pointings.

\begin{figure}
\includegraphics[height=245pt,angle=270]{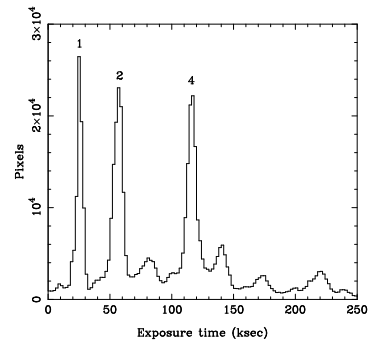}
\caption{{\it Top:} Histogram of the exposure times in all observations used in this work. The narrow peaks lie at the 1, 2, and 4 exposure values generated from the overlapping gridded observations. Peaks are also visible at $\sim$80~ks and $\sim$140~ks where, respectively, 3 and 5 grid pointings overlap due to variations in the roll angles of neighboring ObsIDs, as well as at $\sim$170 and $\sim$220~ks where the existing 50 and 100~ks observations overlap with the central area of 4 grid pointings. The broader bases correspond to overlaps caused by slight variations in the roll angles of the ObsIDs.}
\label{exphisto}
\end{figure}

Figure~\ref{exphisto} shows a histogram of the exposure times, the narrow peaks representing the 1, 2, and 4 ObsID exposure values. Also visible are a number of smaller peaks where the existing 50 and 100~ks observations overlap with the deep central region (peaks at 170 and 220~ks) and where variations in the roll angles of neighboring ObsIDs lead to areas covered by 3 and 5 grid pointings (peaks at 80 and 140~ks respectively). Based on the cumulative fraction of exposure times, approximately 95\% of the 0.97~deg$^2$ survey area has an exposure of at least 30~ks, $\sim$70\% has at least 60~ks, and $\sim$40\% has $\sim$120~ks. 25\% of the survey area exceeds 120~ks due to offset roll angles and the two deeper pointings.

\subsection{Removal of Cyg~X-3 Readout Streaks}
\label{s-cygx3}

\begin{figure}
\includegraphics[width=245pt]{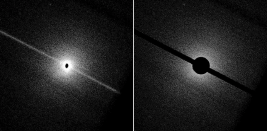}
\caption{Example image of the Cyg~X-3 readout streak in ObsID 10964 before ({\it left}) and after ({\it right}) application of a hard mask to remove events from the event file. Each image is approximately 5\arcm\ wide and aligned with north up and east to the left. The readout streak is orientated along a chip column, the position angle of which is dictated by the observation roll angle.}
\label{masking_example}
\end{figure}

When ACIS reads out it is still taking data, and therefore bright sources continue to expose the entire column in which the source lies, producing {\it readout streaks}. Figure~\ref{masking_example} shows the readout streaks caused by Cyg~X-3 that are present in 6 of our ObsIDs (including ObsID~10939 that was split in two). In these images Cyg~X-3 can be clearly seen as the bright source, with the core of Cyg~X-3 faint due to pileup (when two or more photons are detected as a single event often with a bad grade and the photon counting rate is therefore underestimated) and bad event grades (from piled-up photons). When the data are processed, some of the readout streak events are rejected because of bad grades, but the majority remain.

There are a number of reasons why these readout streaks should be removed prior to data analysis. These events are often falsely identified as real sources by source detection codes, though the positions of these sources allow them to be easily removed. More importantly, these events can contribute to the background regions of nearby sources, leading to incorrectly high background estimates and low source significances.

A number of methods were considered for dealing with these readout streaks. Some studies have incorporated the readout streaks into the background model \citep[e.g.,][]{evan10}, though this can be complex when different pieces of software are used for data reduction and analysis. The CIAO tool {\sc acisreadcorr} can be used to remove readout steaks whilst retaining true background photons by using a background spectrum to separate background and readout events. However, because Cyg~X-3 is so bright the background surrounding the readout streaks is actually the wings of the PSF (see Figure~\ref{masking_example}) and therefore has the same spectral shape and this tool cannot distinguish between the two. Another method is to use the CIAO tool {\sc dmfilth}, which replaces pixel values in a source region with events interpolated from surrounding background regions. However the tool does not provide full event information on all reproduced events and so many extraction or analysis tools cannot work on such data products.

The method that was used was to implement a `hard' mask for the data, completely removing all events that fall within certain regions. The exposure map was also masked in this way so that the various data reduction tools used consider these regions to have zero exposure. This was implemented using two masks, one was circular, centered on Cyg~X-3 with a radius of 40~pixels, removing the detailed substructure of {\it Chandra}'s PSF that is evident in the brightest of sources. The second mask is a long rectangle that covers the readout streak along the length of the CCD on which Cyg~X-3 was imaged, with a width of 10--20 pixels (depending on the width of the streak, which is determined by the size of the Cyg~X-3 PSF and therefore the off-axis angle of Cyg~X-3). Both of these distances were estimated as the distances at which the count rates approached that of the background level. These masks were first applied to the exposure map using {\sc dmcopy} and then {\sc dmimgpick} was used to remove events from the level~2 event file that fell within zero-valued pixels of the exposure map. This was successful in removing the readout streaks in all 6 ObsIDs that included Cyg~X-3. An example of this is shown in Figure~\ref{masking_example}.

\begin{figure*}[t]
\includegraphics[width=512pt]{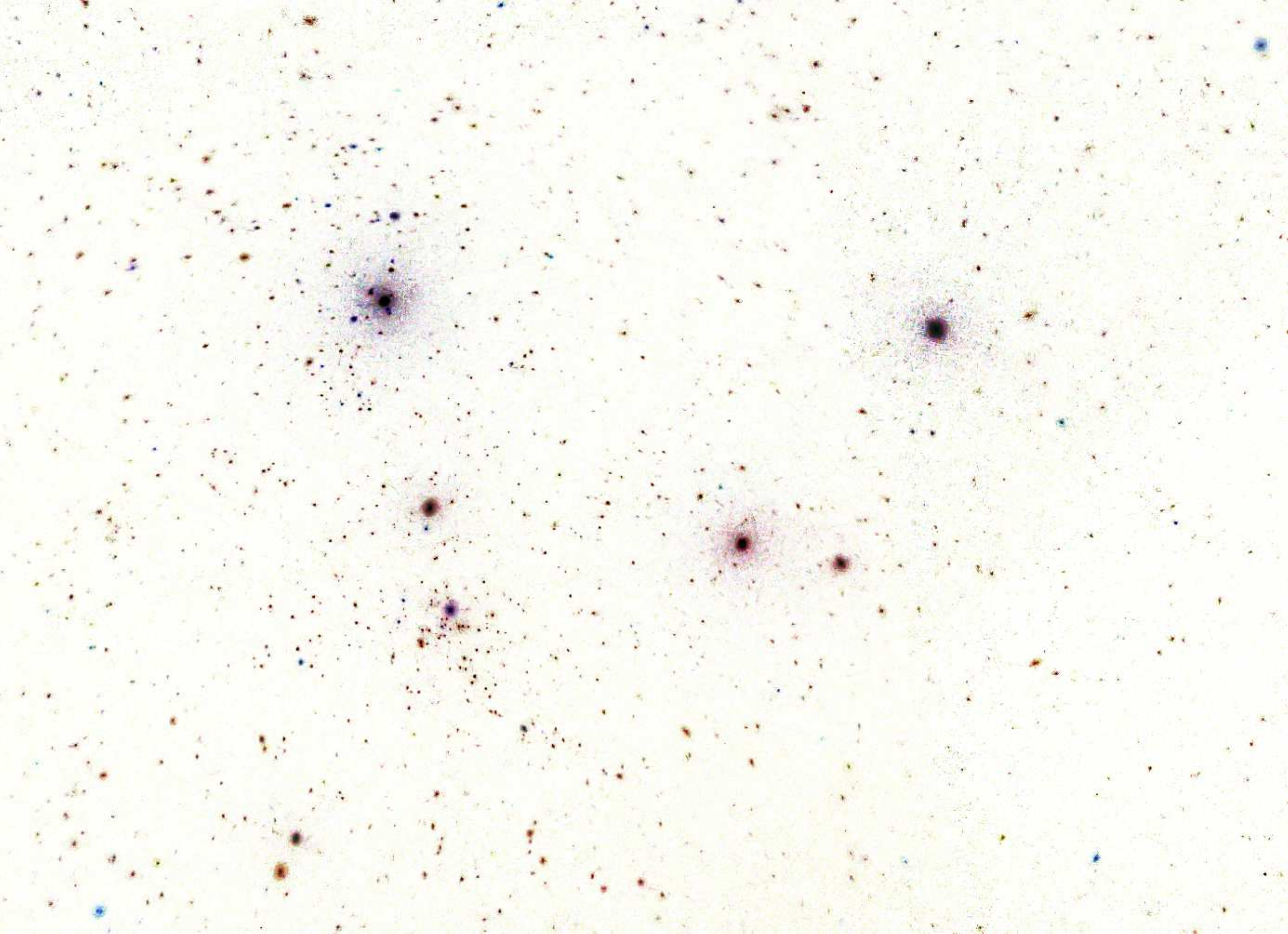}
\caption{An inverted three-color (RGB = {\it soft} 0.5--1.2~keV / {\it medium} 1.2--2.0~keV / {\it hard} 2--7~keV) X-ray image of the center of Cyg OB2 showing observations from the Cygnus~OB2 Chandra Legacy Survey. The X-ray data has been processed, cleansed and flux calibrated as described in the text. The image is approximately $23^\prime \times 17^\prime$ with North up and East to the left. The brightest X-ray sources visible in this image are predominantly OB stars, with the six most luminous objects in the centre of the image being the trapezium of Cyg~OB2 \#8 (upper left), \#9 and \#22 (center and lower left), the blue hypergiant Cyg~OB2 \#12 and MT267 (center right) and Cyg~OB2 \#5 (upper right). See \citet{negu08} for a near-IR image with a similar field of view. Many hundreds of lower-mass stellar X-ray sources are also visible.}
\label{3color}
\end{figure*}

Figure~\ref{3color} shows an inverted three-color X-ray image of the center of the association made by combining the completely processed and cleansed X-ray data in the {\it soft}, {\it medium} and {\it hard} bands. The image has been mosaiced using CIAO and flux calibrated by dividing by the exposure map. The figure reveals the large number of resolved X-ray point sources visible in the data.

\section{Source Detection}

In this section we outline the methods employed to detect possible X-ray sources using a variety of tools. The validity of these sources is determined using more complex tools that take into account the source and background photons in multiple ObsIDs, resulting in a statistical quantification of the source validity (see Section~\ref{s-extraction}). Therefore the goal of source detection is to detect as many sources as is feasibly possible such that after applying an iterative and stringent source validation program the majority of acceptable sources are retained and the weak or false sources weeded out. This method is adopted over an initially conservative source detection method as, despite the extra work involved, we believe it is more likely to produce a larger list of valid sources.

In order to fully exploit the wide and deep observations, and the fact that the majority of sources would be detected in multiple ObsIDs at different off-axis angles, particular care had to be taken to maximize the number of sources that could be detected. We applied three different source detection algorithms at a range of spatial scales and in several different energy bands, specifically the CIAO {\sc wavdetect} \citep{free02} and Palermo Wavelet Detection \citep[PWDetect,][]{dami97} codes, as well as a new and enhanced multi-ObsID version of CIAO {\sc wavdetect}. We adopted three energy bands for the source detection: {\it soft} (0.5--2.0~keV), {\it hard} (2.0--7.0~keV), and {\it broad} (0.5--7.0~keV). We used an upper limit of 7~keV instead of the more typical value of 8~keV because above this energy there is a rise in the instrumental background due to charged particle impact, combined with fluorescent instrumental lines of Ni~K and Au~L (see the POG), which reduces the detection significances of faint sources. To maximize the number of sources in our final catalog we complemented our candidate source list with the positions of known sources in Cyg~OB2, visually inspecting all sources in the X-ray image before adding them.

\subsection{{\sc wavdetect}}

The CIAO tool {\sc wavdetect} was run on each CCD of each ObsID using wavelet scales of 2, 4, 8, 16, and 32 pixels (to be sensitive to both point-like and moderately extended sources, or sources at large off-axis angles) and at multiple detection thresholds. Our first list of candidate sources was generated using a conservative threshold of $S_0 = 10^{-6}$ \citep[see][]{free02}, and then supplemented by additional lists generated using liberal thresholds of $S_0 = 10^{-5}$ and $10^{-4}$. The highest threshold at which each source was detected was stored for later reference when the source lists were merged. Whilst a significance threshold of $10^{-4}$ is rarely used (due to the large fraction of spurious sources it generates) it was used in our work because of the need to detect faint sources observed in multiple ObsIDs and at different off-axis angles. These sources were only retained if they were detected in another ObsID or with another method. We found that across our observations the number of sources detected at thresholds of $10^{-6}$:$10^{-5}$:$10^{-4}$ was approximately in the ratio 1:1.4:2.6.

\subsection{PWDetect}

Source detection was also performed with the wavelet-based source detection algorithm {\it PWDetect} \citep{dami97} at a detection limit of $\sigma = 4.1$, which should produce $\sim$20 spurious detections within each CCD that the code is run on. PWDetect is optimized to take into account the spatial variation of the PSF across the {\it Chandra} FoV, such that at a given position the wavelet transform is only made at scales no smaller than about 1/2 the PSF sigma (since there is no useful information smaller than this).

\subsection{{\it Enhanced} {\sc wavdetect}}

A limitation of the two source detection algorithms used above is that they cannot be run on multiple observations with different field centers. Non-aligned X-ray observations cannot be merged in the traditional sense because the PSF size and shape varies as a function of the off-axis angle, and therefore traditional wavelet-based source detection that takes into account the specific PSF shape at a given point on the CCD would be misguided. Our tiled observing strategy means that for a typical constant source a single ObsID only accounts for 25\% of the observing time, thus limiting our potential to detect the faintest sources possible. 

To overcome this we have developed an enhanced version of CIAO's {\sc wavdetect} that allows source detection on multiple overlapping observations. This improves the sensitivity of the survey and allows us to detect weak sources in the full dataset that are below the detection threshold of individual observations. The method uses the two components of {\sc wavdetect}, {\sc wtransform} and {\sc wrecon}, to self-consistently take into account the size of the PSF in different observations. The {\sc wtransform} routine, which carries out a wavelet transformation of the data, is applied to all observations at the appropriate PSF size. Since wavelet transformation is a linear process, the correlation value of a wavelet scale applied to a combination of datasets is simply $C_{ij}^{(1+2+...)} = C_{ij}^{(1)} + C_{ij}^{(2)} + ...$ (as long as the PSF size criteria is met for all datasets). Only observations where the wavelet scale is larger than the PSF size are combined, thus limiting the number of false detections that arise at scales smaller than the PSF size. We recompute the detection threshold at each pixel for each scale and detect a source if $C_{ij}^{i+j+...} > \mathcal{T}_{ij}^{i+j+...}$. The threshold is computed based on the summed backgrounds from all the observations, $b^{i+j+...} = b^i + b^j + ...$, excluding those observations where the wavelet scale is smaller than the local PSF size, and is therefore equivalent to the threshold used with CIAO {\sc wavdetect}. Furthermore, because {\sc wrecon} is used exactly as it would be with CIAO's {\sc wavdetect}, both the thresholds and the number of false positives remain the same.

We ran the code on ``tiles'', which consist of multiple overlapping CCDs built up from our tiled observing strategy. Our observations consisted of 49 tiles, of which 45 have two or more observations overlapping. The {\it Enhanced} {\sc wavdetect} procedure was therefore run on each tile + band combination. We used a threshold of $S_0 = 10^{-5}$ (more conservative than used with {\sc wavdetect} because of the large number of sources that were detected at more liberal levels of detection) which resulted in a total of 21,121 detections. The vast majority of these were detections of the same source at different spatial scales and in different energy bands, or were duplications of sources already detected by {\sc wavdetect}. These duplications were removed following the same technique as employed by {\sc wavdetect} itself, leaving a total of 2,635 new candidate sources from running the {\it Enhanced} {\sc wavdetect}.

\subsection{Additional lists of known sources}

The source detection methods outlined above have hopefully detected the majority of significant X-ray sources in our observations, but given the complex nature of both X-ray observations and our tiled observing strategy it is possible some faint sources may have evaded detection. Lowering the source detection threshold would likely increase the number of valid sources detected (those that would survive our source verification process) but would also have dramatically increased the number of false sources detected (those that would be discarded at the source verification stage). Faced with a situation of diminishing returns it is natural to draw a line in the source detection process at a certain level. However, some valid source may have escaped detection and thus other methods that could identify these sources should be considered. Whilst the primary objective of these X-ray observations is to compile a catalog of previously unknown members of Cyg~OB2 there is also significant scientific merit in studying the X-ray properties of previously known members of the association.

For this reason we supplemented the list of X-ray detected sources with 634 additional sources from lists of known Cyg~OB2 members, including O and B-type stars \citep{wrig14c}, young A-type stars in the association \citep{drew08}, and lower mass stars at unique evolutionary phases \citep{wrig12a}. These sources had not been detected in the original source detection methods outlined above but were included in the list of candidate X-ray sources to be verified in our full extraction process. Following the full source extraction and verification process only 100 of these sources were retained in our catalog, and these are noted in the final catalog. These 100 sources comprise $\sim$1\% of the final catalog.

\subsection{Combining the source lists}

The different source detection methods outlined above produced a total of 54,311 sources (22,443 from {\sc wavdetect}, 28,599 from {\it PWDetect}, 2,635 additional sources from the {\it Enhanced} {\sc wavdetect}, and 634 additional sources), many of which were duplicates detected by different methods and therefore needed to be removed. To understand the extent of the source duplication we performed a simple comparison of the source detection results from the main two codes, {\sc wav detect} and {\it PWDetect} (which were responsible for the vast majority of sources detected), on a single `tile' of four overlapping CCDs in our observations (after merging the multiple source lists from the four overlapping CCDs and removing multiple detections).

For this experiment {\sc wavdetect} was run at a threshold of $S_0 = 10^{-6}$ and {\it PWDetect} was run with a detection limit of $\sigma = 4.7$, both of which should statistically result in 1 false source per CCD (therefore 4 false sources per `tile'; note that these false source numbers are estimates and may not be correct), {\sc wavdetect} detected 148 sources and {\it PWDetect} 185 sources with an overlap of 125 (overlap fractions of 84\% and 68\% respectively). Decreasing the source detection thresholds for both algorithms (and thereby increasing the number of expected false sources to $\sim$10 per CCD - though there may be more or less than this) resulted in overlap fractions of 71\% and 57\% for the two methods (detection increases of 39\% and 38\% respectively), suggesting there was an increase of un-matched `false positives' and a tendency for the two algorithms to uncover different faint sources. {\it PWDetect} generally detected more sources in our tests, particularly at large off-axis angles, but the positions it determined were less accurate (with a larger standard deviation between positions detected from different ObsIDs and positions of known sources). Sometimes a source would be identified as one source by one code, but as two sources by another code (it was more common for {\sc wavdetect} to identify a source as two separate sources). In these situations it was our belief that both sources should be taken through to the source verification stage and assessed there.

Because of the larger number of detections and the high overlap between source detection methods an automated detection merging process was required. The method employed was based on that used by the {\it Chandra} Source Catalog \citep[][see Appendix~A]{evan10} and begins by building `groups' of detections with overlapping error ellipses. This method is based on the assumption that all our sources are point-like (a valid assumption given that the great majority of sources are expected to be stellar) and that merging detections at different energies is no different from merging detections from different source detection algorithms.

Three types of group are possible from this method: unambiguous single detections, unambiguous multiple detections, and ambiguous multiple detections. The first of these is a single detection that does not overlap with any others and so is preserved as a single `source' (there were 8568 of these in our detection lists). The second of these is a group of multiple detections all of whose error ellipses overlap with all the other error ellipses of detections in the group. This group is therefore consistent with being a single source, of which we find 3124 in our detection lists. Finally an ambiguous group of multiple detections consists of detections whose error ellipses overlap, but not all error ellipses overlap with all other error ellipses and is therefore not consistent with being a single source. Visual inspection showed that the vast majority of these (of which there were 1349 in our detection list) were caused by multiple close sources whose error ellipses overlapped in one or more detections leading to a `chain' of detections (with group sizes of between 3--32 detections). These groups were resolved by visual inspection, adopting detections and error ellipses made at the smallest off-axis angle (and therefore the best spatial resolution) in uncertain cases. Finally, as described in Section~\ref{s-extraction}, all sources were visually inspected to remove obviously spurious sources due to detector artifacts such as readout streaks.

\subsection{Final candidate source list}

\begin{figure*}
\begin{center}
\includegraphics[width=440pt, angle=270]{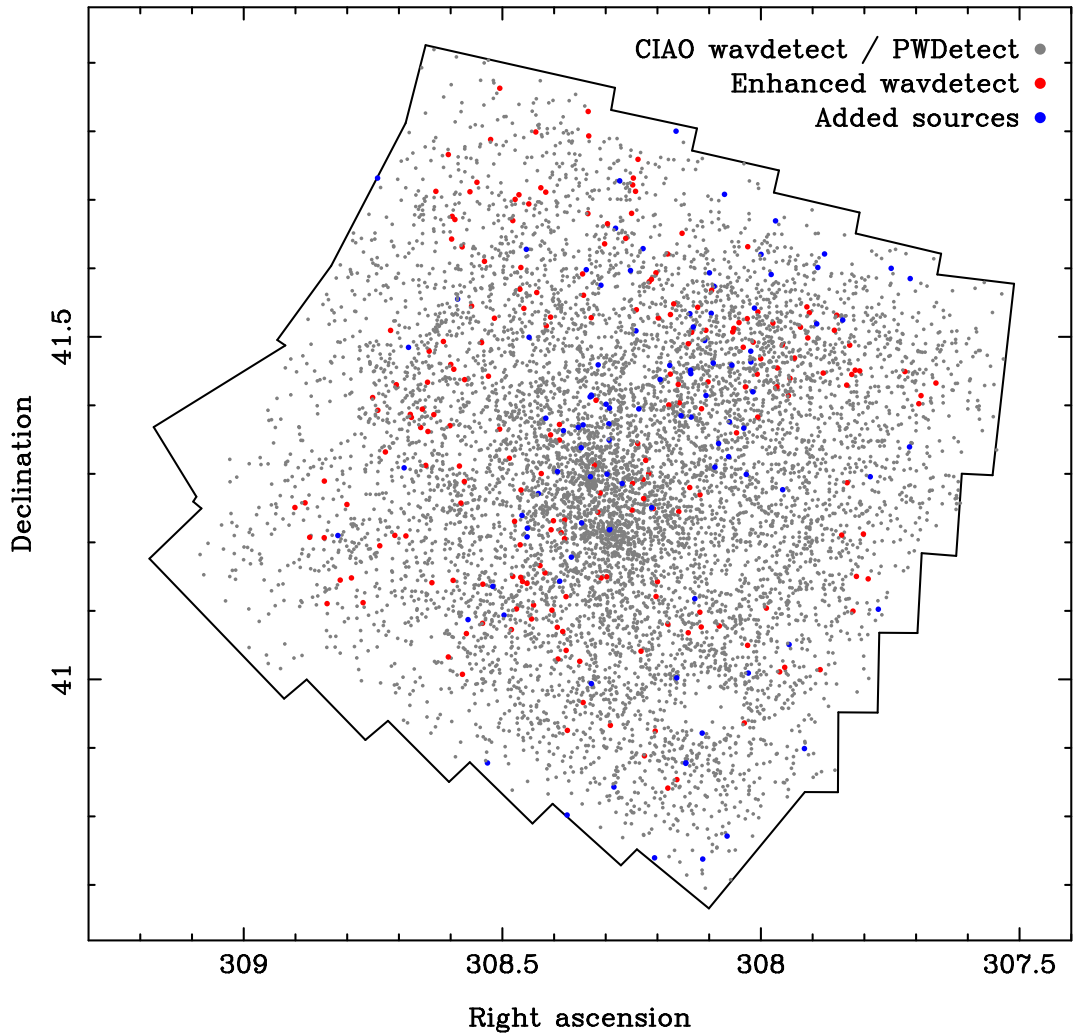}
\caption{Final distribution of the 7924 X-ray sources in the {\it Chandra} Cygnus OB2 Legacy Survey after removing all sources that failed the source verification procedure (Section~\ref{s-extraction}). The sources are color-coded based on the detection method used to add the sources to the initial source candidate list. The 7608 surviving sources detected using either {\sc wavdetect} or {\it PWDetect} (mostly detected using both codes) are colored grey. The 216 surviving sources detected using the {\it Enhanced} {\sc wavdetect} code (but not detected by other methods) are colored red and the 100 surviving sources added from known source lists (and not detected by other methods) are colored blue. \vspace{0.0cm}}
\label{detection_map}
\end{center}
\end{figure*}

This process led to a final total number of 13,041 groups which would hereafter be treated as, and referred to as, candidate sources. The positions of the final sources were calculated based on the most on-axis detections of a group (note that positions would be revised later during the full photon extraction). Figure~\ref{detection_map} shows the spatial distribution of the final 7924 sources after the full source verification process, color-coded by their detection method.

\section{Point Source Extraction}
\label{s-extraction}

In this section we describe the process of extracting point sources from the data, including assessing the validity of the sources, refining our source list, and adjusting the positions of our sources. Point source extraction was performed using the {\em ACIS Extract}\footnote{The {\em ACIS Extract} software package and User's Guide are available online at \url{http://www.astro.psu.edu/xray/docs/TARA/ae\_users\_guide.html}.} \citep[AE,][]{broo02,broo10} software package, which defines source extraction apertures, extracts source and background spectra, and compiles source photometry, spectra, and lightcurves.

Our source extraction process is divided into two stages, an iterative source verification process (in which candidate sources are extracted and their validity assessed) followed by a full spectral extraction (in which the properties of the source are to be accurately determined). Because the objectives of these two stages are different and, because for the majority of our candidate sources we have multiple observations available to us, we use different combinations of the data to achieve these goals. In the first stage of the extraction, when we wish to assess the validity of our proposed sources, we use the subset of whole observations that optimizes (or minimizes) the quantity $P_B$, the probability that the source is a background fluctuation. This is based on the methodology that a source is deemed to exist if it is significant in any observation, or in any combination of observations. In the second stage we wish to maximize the signal to noise ratio (S/N) of any measured quantities through our extraction. In this instance AE uses a greater subset of the observations, discarding data only when retaining them would significantly degrade the final S/N. These differences are most often manifested through discarding or retaining data at large off-axis angles when data at smaller off-axis angles also exists for a source. A full discussion of this approach can be found in Section~6.2 of \citet{broo10}.

Because this is the fundamental difference between the two stages of our extraction process we first outline the general extraction process used by both stages, and then highlight the important parts of the extraction process relevant to the two individual stages. A full description of the AE package can be found in \citet{broo10}, but we reiterate the main steps here.

\subsection{The extraction process}

A basic extraction process is used whether the goal of the extraction is to assess source validity, determine accurate source positions, or extract source properties. This process begins with assigning an extraction aperture for the source (Section~\ref{s-apertures}) based on the local PSF and level of crowding. Then a background region is defined (Section~\ref{s-background_regions}) so as to accurately characterize the background contamination in the source aperture. Events are then extracted from these two regions (Section~\ref{s-source_extraction}) and the results compared with a model of the source and background that includes the properties of the {\it Chandra} observatory so that the intrinsic source properties can be calculated. These steps are described here.

\subsubsection{Extraction apertures}
\label{s-apertures}

Because the {\it Chandra} point-spread function (PSF) is both non-circular and also varies significantly across the field of view, the extraction of point-like sources requires a model of the local PSF for each observation of each source (because a source may be observed on-axis in one ObsID and off-axis in another, causing the PSF to vary significantly). AE builds finite extraction apertures from contours of the local PSF at an energy of 1.5~keV enclosing, by default, 90\% of the PSF power, decreasing this fraction (to a minimum of 40\%) in crowded regions to prevent overlapping extraction apertures from different sources. If a close pair of sources has their extraction apertures decreased to less than 40\% of the PSF power then either that observation is not considered by AE for these sources (when multiple observations are available), or else the weaker of the two sources is discarded (on the belief that no meaningful source properties can be determined for the source). In this situation the extraction aperture of the stronger source will increase to the default value of 90\%, encompassing the weaker source. This situation therefore reverts to the default situation of a single source, which is appropriate if one cannot reliably prove there to be two sources present.

A hook-shaped feature, $0^{\prime\prime}.8$ in length, was discovered in the {\it Chandra} PSF in 2011,\footnote{http://cxc.harvard.edu/ciao/caveats/psf\_artifact.html} extending from the peak of the source in a roll-angle dependent position. It is estimated to contain $\sim$5\% of the source flux and is therefore only discernible for bright sources observed sufficiently on-axis that the PSF and the hook are compact enough to be individually identified. At the time of writing the exact flux fraction, shape and energy dependence of the feature have not been well characterized and no revised PSF models have been produced. It is therefore possible that our source detection and extraction process may have detected this `hook' as a real source and verified its authenticity. AE includes a tool that identifies where bright PSF hooks would appear in the available data, facilitating an inspection of all sources bright enough ($>$100 net counts) and observed on-axis ($\leq$4$^\prime$) such that the hook might be identified as a separate source. We inspected 688 sources that met these two criteria but could not identify any sources that appeared consistent with being due to the PSF hook feature. The visual inspection process was aided by the fact that each source was typically observed at multiple off-axis angles and roll angles due to our tiling strategy. Since the PSF hook feature would therefore have appeared in different positions in each observation it was straightforward to verify that none of our extracted sources were due to the PSF hook.

\subsubsection{Background regions}
\label{s-background_regions}

Background estimates must be obtained ``locally'' for each source to account for spatial variations in the X-ray background due to diffuse emission or the wings of nearby point sources. For an isolated source a simple background annulus is used, extending from a radius 1.1 times that which encloses 99\% of the PSF to a radius such that a minimum of 100 background photons has been gathered, excluding regions around other point sources. AE also adjusts the sizes of the background regions used so that the Poisson uncertainty on the background level contributes no more than 3\% of the total uncertainty on the source photometry.

In crowded regions it is often not possible to construct background regions that avoid other point sources, especially since the PSF wings of bright neighbors can be especially wide. To overcome this we employ AE's alternative background algorithm that constructs background regions that sample nearby bright sources in proportion to their expected contamination of the extraction aperture of the source in question. This is estimated using a spatial background model for all the nearby sources determined using PSF models and estimates of their fluxes, and is then improved over multiple iterations and extractions. 

\subsubsection{Source extraction}
\label{s-source_extraction}

Once the extraction apertures are defined for each source, events are extracted by AE using standard CIAO tools, including {\sc dmextract} to construct source spectra, {\sc mkarf} to build ancillary reference files (ARFs), and {\sc mkacisrmf} to construct response matrix files (RMFs). Corrections are then applied for the finite size of the PSF and the lost source events falling outside of the extraction aperture. Because {\it Chandra}'s PSF is energy dependent this correction is calculated at multiple energies and convolved with the ARF to represent the true observatory effective area responsible for the extracted source and background events.

\subsubsection{Source repositioning}
\label{s-positions}

During the extraction process the positions of source candidates, originally derived from wavelet-based source detection, are updated with AE estimates using a subset of each source's extractions chosen to minimize the positional uncertainty. AE offers three positional estimates: the {\it data} position, calculated from the centroid of all extracted events; the {\it correlation} position, derived from the spatial correlation between the {\it Chandra} PSF and the events; and the {\it maximum-likelihood} position, based on the peak in the maximum-likelihood image reconstruction of the source neighborhood. For uncrowded on-axis sources these positions are very similar and therefore the {\it data} position is the easiest to calculate. For off-axis sources the asymmetry in the {\it Chandra} PSF can bias this position so the {\it correlation} position is often more accurate, while for crowded sources with overlapping PSFs the {\it maximum-likelihood} position provides better positional estimates. We follow AE's guidelines by using this system to allocate positions for all sources while also visually inspecting all repositioned sources. Source repositioning was performed every 2--3 iterations (see Section~\ref{s-trimming}) to provide accurate source positions for assessing the source validity during the extraction process and typically resulted in shifts of $\leq$0.1\arcs.

\subsection{Iterative trimming of the source list}
\label{s-trimming}

Our liberal source detection strategy relies upon a careful and conservative source verification process such that weak point sources likely to be background fluctuations are removed and only those sources that pass a critical threshold are retained. An accurate assessment of the significance of a candidate source requires a full extraction of the source and background regions that is not possible through the simple source detection process, especially for complex observational datasets such as ours, and must be assessed following a full point source extraction. This requires an iterative process of source extraction and verification.

Traditional source validity is assessed via the signal to noise ratio, defined by the source flux divided by its uncertainty, a quantity that is equivalent to a source significance when the source and background fluxes have Gaussian distributions. However, since most X-ray sources in our sample are quite weak and background count rates are low, the Gaussian approximation is not valid and photon distributions follow Poisson statistics. To overcome this AE assesses the significance of a source using the probability that a source can be explained as a background fluctuation according to Poisson statistics \citep[][Section~4.3]{broo10}, equivalent to testing the null hypothesis that a candidate source does not exist \citep[][Appendix~A2]{weis07}. When multiple observations of a source exists AE bases the calculation of $P_B$ on the subset of each source's extractions that maximize the significance of the source. This prevents extractions at large off-axis angles with a large PSF from biasing the measured significance of a source that is also observed on-axis with a small PSF.

The threshold {\it `not-a-source probability'}, $P_B$, that is chosen must balance completeness (the fraction of real sources detected) against reliability (the fraction of sources that are `false'). We follow the standard threshold adopted by previous users of AE and require our sources to have a probability of being a background fluctuation of $P_B \leq 0.01$ and calculated from a minimum of 3 net counts. All sources that do not meet these requirements are reviewed visually and then discarded. The visual review is intended to prevent sources that have been wrongly demoted from being lost, such as sources where the extraction aperture was not correctly centered on the source. Such instances were very rare and the sources were retained for another round of extraction and if necessary their extraction apertures were corrected.

Once these faint sources have been discarded the entire catalog is extracted once more, including recalculating source and background regions since these may change if nearby sources have been removed. The net effect of removing sources is that more events will contribute to the background regions of nearby sources, thereby lowering their significance. Therefore sources that survive the source pruning may now fall below the source criteria, requiring another round of source pruning. An iterative sequence of alternating source extraction and source pruning is therefore required and continues until no sources are found to be insignificant. To prevent the accidental removal of a large number of potentially valid sources in an early iteration we started this process with a probability threshold higher ($P_B = 0.1$) than our final threshold, adopting a threshold of $P_B = 0.01$ after three iterations. This process left a list of 7924 valid sources. At this stage the remaining sources are considered to be valid sources and a full spectral extraction can be performed.

\subsection{Full Spectral Extraction}
\label{s-fullextraction}

Once a high-confidence list of sources has been produced we perform a full spectral extraction on the sources, following the basic steps outlined above, but this time choosing a subset of the data that optimizes the photometric S/N of any measured quantity such as source photometry or spectroscopy, but without succumbing to photometric bias. AE achieves this by discarding observations only when retaining them would halve the maximum S/N that could be achieved with the observations. This strategy tolerates a slight deterioration to the S/N in order to avoid photometric biases arising from data selection \citep{broo10}.

Source spectra are extracted along with background spectra and facilitate the calculation of many observed quantities such as the number of source, background, and net photon events, the photon flux (in photons cm$^{-2}$ s$^{-1}$) and the photon energy flux (in erg cm$^{-2}$ s$^{-1}$). Additional quantities with a high diagnostic value include background-corrected quartiles (25\%, 50\%, and 75\%) of the observed event energy distributions (which provide a valuable characteristic of the observed spectrum for low-count data) and the probability that the photon arrival times can be described by a source with a constant flux (a useful diagnostic of variable sources and an indicator of flare-like events).

\subsection{Source Variability}
\label{s-variability}

To quantify the level of X-ray variability of a source we follow \citet{broo10} and use a one-sample Kolmogorov-Smirnov (KS) test, as implemented by AE, comparing the arrival times of events (in calibrated units of photon ks$^{-1}$ cm$^{-2}$) with the null hypothesis of a uniform source flux. This test then provides a probability (a $p$-value) that the source is not variable and is assessed both within individual observations and between all observations (accounting for variations in effective area among the observations). While this test does not reveal periodicity or variability near the beginning or end of the observation, and does not take into account variability in the background, it does provide a simple indication of potentially variable behavior that is useful to identify which sources are worthy of further investigation.

This variability will be discussed in more detail in \citet{flac14b}, but we briefly summarise the single-parameter results here. We consider a source as being variable if the $p$-value of its light curve being reproduced by a source of constant flux is $\leq 0.005$ ($\sim$3$\sigma$ confidence of variability). We identify 628 (8\%) sources as being variable within a single observation, while 2193 (28\%) exhibit evidence of inter-observation variability (with an overlap between these categories of 432 sources). Because these two variability tests are not independent of each other, the results from the tests should be compared with care. In total, 2389 (30\%) sources exhibit some evidence of variability from this single parameter.

\section{Source catalog properties and statistics}

The final source list contains 7924 X-ray sources that pass our threshold significance criteria and are considered to be valid sources. The properties of these sources are presented in a table published electronically and available at the Vizier archive\footnote{Weblink: http://vizier.u-strasbg.fr/cgi-bin/VizieR}. The column names and descriptions are listed in Table~\ref{columns}.

\input{columns.tex}

\subsection{Technical properties}

\begin{figure*}
\begin{center}
\includegraphics[height=500pt, angle=270]{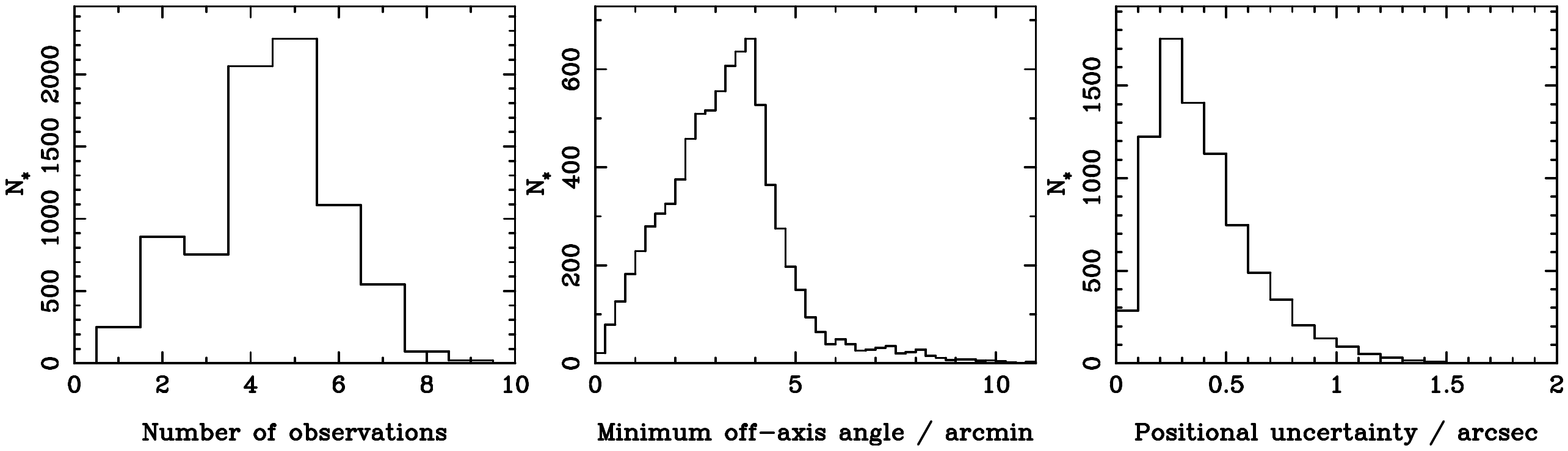}
\caption{Distribution of various observational and technical quantities for the 7924 sources in the final X-ray catalog. {\it Left:} The number of times each source is observed. The majority of sources ($>$80\%) are observed at least 4 times due to the tiling strategy outlined in Section~\ref{s-design}. {\it Center:} The smallest off-axis angle at which any source is observed, with quartiles at 2.3, 3.2, and 4.0 arcmin. {\it Right:} The positional uncertainty of all sources, with quartiles at 0.2, 0.3, and 0.5 arcsec.}
\label{extraction_histograms}
\end{center}
\end{figure*}

\begin{figure*}
\includegraphics[width=440pt, angle=270]{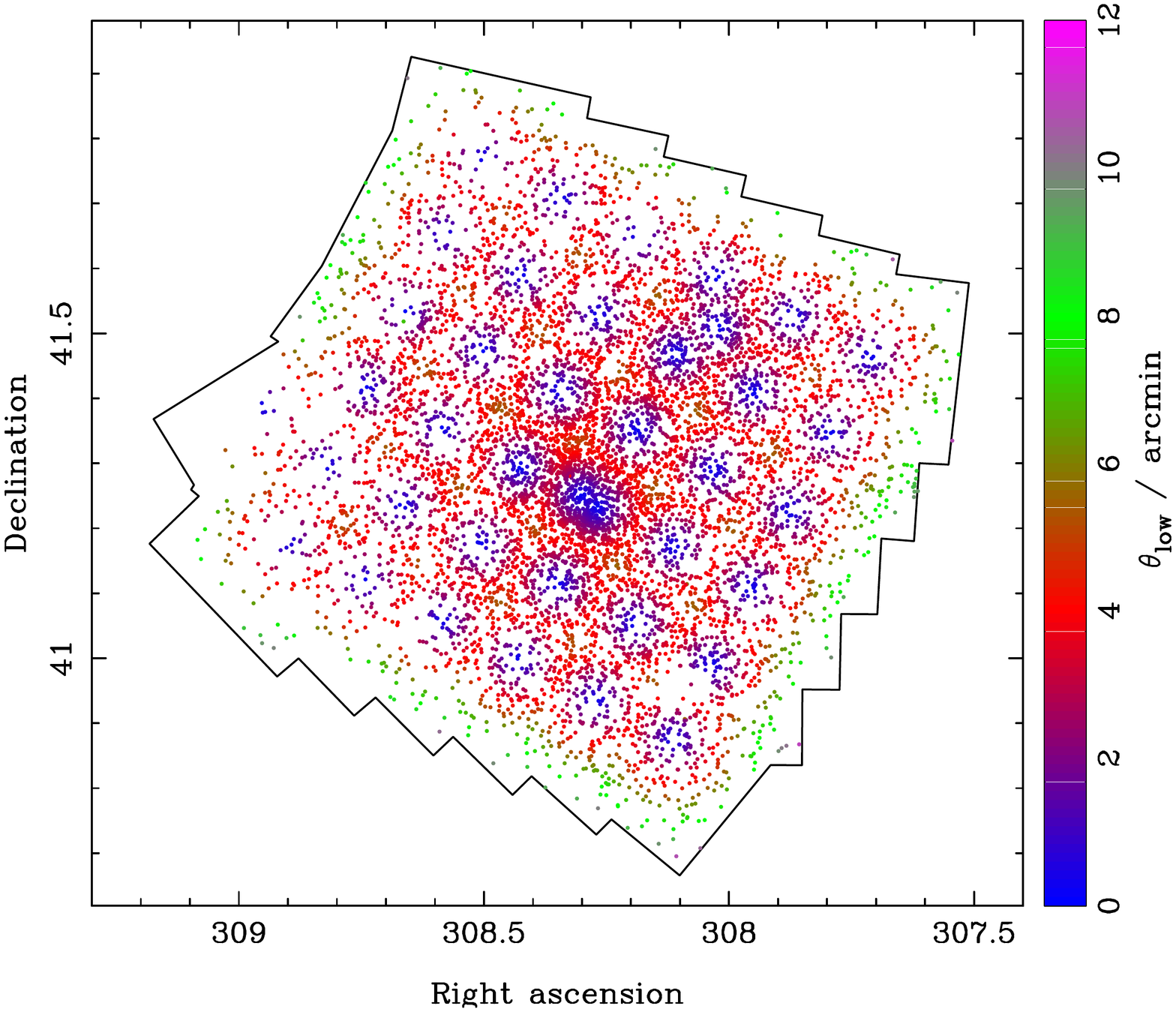}
\caption{Distribution of X-ray sources in the {\it Chandra} Cygnus OB2 Legacy Survey colored as a function of the lowest off-axis angle that the source is detected at. The observational tiling strategy is evident in the grid pattern seen in this image. Due to the tiling strategy used the vast majority of sources were observed at least once at an off-axis angle $<$4$^\prime$ and therefore with PSF sizes $<$2$^{\prime\prime}$ in radius (at 1.49~keV).}
\label{offaxis_map}
\end{figure*}

Figure~\ref{extraction_histograms} shows various observational properties of the final source catalog of 7924 sources. The vast majority of sources are observed more than once thanks to the observational tiling strategy adopted, with $>$80\% of sources observed 4 or more times (see Section~\ref{s-design}). The principle product of the survey design is that the overlapping grid of pointings means that the majority of sources will be observed on-axis at least once, and this is also evident in Figure~\ref{extraction_histograms}, where it can be seen that half of all sources are observed at least once with an off-axis angle of at most 3.2\arcm, and 75\% of sources are observed at least once with an off-axis angle of $<$4.0\arcm. This can be seen in relation to the survey area in Figure~\ref{offaxis_map}, which shows the distribution of all X-ray sources colored by the smallest off-axis angle at which they are observed. This has the effect of highlighting the tiling strategy used for the observations to an extent that is not as evident from the distribution of sources in Figure~\ref{detection_map}. One of the important ways that the off-axis angle translates into the final sources properties is that the positional uncertainty is very low. Figure~\ref{extraction_histograms} shows that the vast majority of sources have a positional uncertainty $<$1\arcs\ and 50\% of sources have an uncertainty less than 0.3\arcs.

\subsection{Source properties}
\label{s-sources}

\begin{figure*}
\begin{center}
\includegraphics[height=500pt, angle=270]{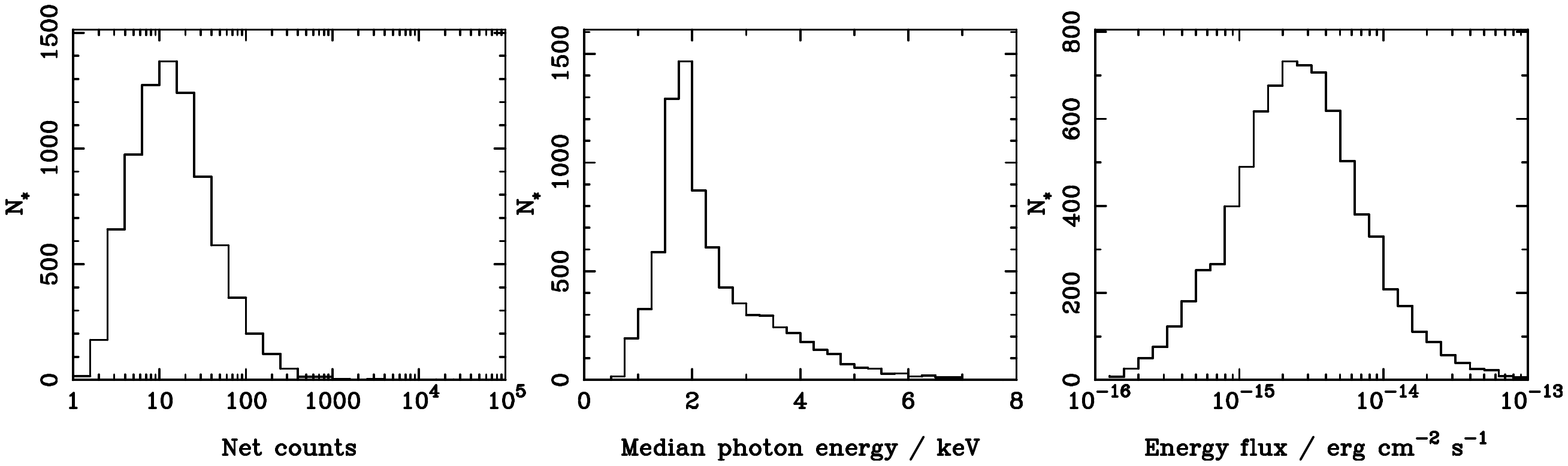}
\caption{Distribution of various source quantities for the 7924 sources in the final X-ray catalog. {\it Left:} The distribution of net counts of all extracted sources, with quartiles at 7, 13, and 28 net counts. {\it Center:} The background-subtracted median photon energy of all extracted sources, with quartiles at 1.7, 2.0, and 2.8 keV. {\it Right:} The distribution of model-independent source energy fluxes calculated using the equation $F_{energy} = 1.602 \times 10^{-9} \tilde{E}_{photon} \times F_{photon}$ (see Section~\ref{s-sources}) with a median of $2.6 \times 10^{-15}$~erg~cm$^{-2}$~s$^{-1}$.}
\label{source_histograms}
\end{center}
\end{figure*}

The distribution of source properties is shown in Figure~\ref{source_histograms} for all 7924 sources in the final X-ray catalog. Most sources are very weak, only 402 sources (5\% of the catalog) have more than 100 net counts, and 39\% have fewer than 10 net counts. Figure~\ref{source_histograms} shows the net count distribution for all the sources in our catalog, extending up to the most luminous X-ray sources with $\sim$60,000 net counts. There are 15 sources with more than 1000 net counts in our sample, the vast majority of which (13 of 15) are known O-type stars, with those having the most net counts being supergiants \citep[for a discussion of the X-ray properties of the known O and B-type stars in Cyg~OB2, see][]{rauw14}. The turnover at low net counts is due to the source verification criteria used, which, it should be noted, is not necessarily based upon the same extracted events as those used to derive source properties. This explains the presence of `verified' sources with apparently $\sim$1--2 net counts in Figure~\ref{source_histograms} (see discussion in Section~\ref{s-extraction}). Note that the intensities of many of the weak sources near the threshold of detectability will be affected by Eddington bias \citep{eddi40} and we warn the user that luminosities estimated directly from the measured counts will tend to be overestimated. Accounting for this requires proper modelling of the Poisson distribution and the detection threshold across the detector.

Figure~\ref{source_histograms} also shows the distribution of background-subtracted median photon energies for all sources in the final catalog. The median photon energy is a commonly-used diagnostic of the X-ray source emission, and for young stellar sources can also be an indicator of the absorbing column density \citep{feig05}. The distribution of median photon energies is strongly peaked at around $\sim$2~keV, in agreement with previous studies of Cyg~OB2 \citep[e.g.,][]{wrig10a}, and slightly higher than the typical peak values for unobscured field stars \citep[$\sim$1~keV, e.g.,][]{wrig10b} and for stars in less-obscured star forming regions \citep[$\sim$1.3~keV, e.g.,][]{getm05a}.

Finally Figure~\ref{source_histograms} shows the distribution of model-independent energy fluxes, which are defined by \citet{broo10} as

\begin{equation}
F_{energy} = 1.602 \times 10^{-9} \tilde{E}_{photon} \times F_{photon}
\end{equation}

where $\tilde{E}_{photon}$ is the background-subtracted median photon energy, $F_{photon}$ is the photon flux \citep[][Section 7.4]{broo10}, and the numerical constant arises from the conversion between keV and ergs to produce units of ergs~cm$^{-2}$~s$^{-1}$. The distribution of energy fluxes rises towards fainter sources and peaks at $\sim 3 \times 10^{-15}$~erg~cm$^{-2}$~s$^{-1}$, which provides a first-order estimate of our completeness limit over the entire survey area (note that the survey completeness limit will be deeper in the central 0.5~deg$^2$ of the survey where the exposure time is higher). The most luminous sources are the OB supergiants Cyg~OB2 \#8A (O6I), Cyg~OB2 \#5 (O7I) and Cyg~OB2 \#12 (B3.5Ia+), which have energy fluxes of $\sim$ 2--4 $\times 10^{-12}$~erg~cm$^{-2}$~s$^{-1}$, though many of these suffer from severe pile-up. No correction was made for pile-up for these sources in our catalog, but we refer the interested reader to \citet{rauw14} who provide a detailed discussion and analysis of pile-up for these sources as part of their analysis of the X-ray properties of the OB stars in Cyg~OB2. Of the ten most luminous X-ray sources, 9 are known OB and Wolf-Rayet stars, though the 9th most luminous X-ray source is a previously unknown and optically faint source that is most likely a flaring pre-MS star.

\subsection{Spurious Sources}

The detection process and source verification process employed by our survey is designed to maximize the detection of all sources, and therefore the detection of a small number of false sources should be expected. To quantify the number of spurious detections one would ideally perform source detection and verification on synthetic observations that reproduce the observations as closely as possible. Variations in the background level may reach such levels as to be detected and characterised as valid sources, and the frequency of these sources could then be quantified. {\bf For example such simulations were performed by the XMM-Newton serendipitous survey \citep{wats09} and the Swift Point Source Catalog \citep{evan14}, both of which found large differences between the false positive rate expected from formal statistics and that found from simulations. However,} due to the complexity of our survey, the multiple, heavily overlapping observations, and the multi-stage detection, inspection and verification process used in producing our source catalog, it would be impossible to perform simulations of our survey in a reasonable amount of time. An alternative approximate estimate of the number of false sources in our catalog may be calculated from the sum of all {\it not-a-source probabilities}, $\Sigma \, P_B$. From our final source catalog we calculate this value to be 5.6 false sources within our catalog, or 0.07\% of the total catalog.

\section{Summary}

In this paper we have presented and discussed the X-ray observations from the {\it Chandra} Cygnus~OB2 Legacy Survey \citep{drak14}, as well as a small number of existing X-ray observations that cover the same region. We have described the standard data reduction, X-ray source detection, and source verification procedures that we have followed. The tiling pattern adopted by our survey leads to a large area at which the vast majority of sources are observed on-axis at least once, including a deep area of 0.5~deg$^2$ with a highly uniform exposure of $116.3 \pm 0.7$~ks, a standard deviation of only 0.6\%. These are pivotal requirements for obtaining high and uniform sensitivity levels over the majority of the survey area and for limiting spatial biases for the analysis of this unique and interesting target. A number of novel data analysis techniques were introduced to maximise the scientific return of our unique set of observations, including an enhanced version of the CIAO tool {\sc wavdetect} that performs source detection on multiple non-aligned X-ray observations, detecting sources that may not be detected in the individual observations.

We have presented a catalog of 7924 X-ray point sources detected and verified from these observations. The catalog, available online, contains positions and basic X-ray properties. The vast majority of sources are observed at least 4 times, and detected on axis ($< 4$\arcm) at least once. The positional uncertainty is therefore very low (typically $< 0.5$\arcs). An analysis of the sensitivity of our survey to a number of observational and stellar parameters is presented in \citet{wrig14e}. Optical and infrared counterparts to these sources are presented in \citet{guar14}, and an analysis of the likely Cyg~OB2 members and contaminants is presented in \citet{kash14}. X-ray spectroscopy of the low- \citep{flac14a} and high-mass \citep{rauw14} members of the association is presented elsewhere, along with analysis and discussion.

\acknowledgments

We would like to extend special thanks to Ken Glotfelty, Frank Primini, and especially Pat Broos for their advice and assistance in this work. We would also like to thank the anonymous referee for a prompt and helpful report that has improved the content and clarity of this paper. NJW acknowledges a Royal Astronomical Society Research Fellowship. NJW and MG acknowledge support from {\it Chandra} grant GO0-11040X. JJD was supported by NASA contract NAS8-03060 to the {\it Chandra X-ray Center} and thanks the director, H. Tananbaum, and the science team for continuing support and advice. This research has made use of data from the {\it Chandra X-ray Observatory} and software provided by the {\it Chandra X-ray Center} in the application packages CIAO and Sherpa, and from Penn State for the {\sc ACIS Extract} software package. This research has also made use of NASA's Astrophysics Data System and the Simbad and VizieR databases, operated at CDS, Strasbourg, France.

\bibliographystyle{apj}
\bibliography{/Users/nwright/Documents/Work/tex_papers/bibliography}

\end{document}

%% file: obsid_table.tex
\begin{table*}
\begin{center}
\small
\caption{Log of {\it Chandra} observations} 
\label{obsids} 
\begin{tabular}{ccccrcc}
\tableline 
Obs. ID 	& Grid	& R.A.	& Dec.	& Roll	& Start Time	& Exp. Time \\ 
 		& number	& (J2000)	& (J2000)	& (deg)	& (UT) 		& (ks) \\ 
\tableline 
4358 & - & 20 32 05.54 & +41 30 31.98 & 195.4 & 2002 Aug 11, 19:50 & 5.0 \\
4511 & - & 20 33 12.22 & +41 15 00.68 & 349.0 & 2004 Jan 16, 10:51 & 97.8 \\ 
4501 & - & 20 32 05.75 & +41 30 38.98 & 170.4 & 2004 Jul 19, 02:03 & 49.4 \\ 
7426 & - & 20 35 47.73 & +41 23 12.98 & 58.4 & 2007 Mar 17, 16:43 & 20.1 \\
10939 & 1-2 & 20 32 06.48 & +40 59 12.30 & 359.8 & 2010 Jan 25, 13:14 & 24.4 \\ 
10940 & 1-3 & 20 31 47.00 & +41 06 19.22 & 1.0 & 2010 Jan 26, 12:05 & 28.1 \\ 
10941 & 1-4 & 20 31 27.53 & +41 13 26.14 & 2.6 & 2010 Jan 27, 19:00 & 30.1 \\ 
10942 & 1-5 & 20 31 08.05 & +41 20 33.06 & 4.2 & 2010 Jan 29, 03:00 & 27.3 \\ 
10943 & 1-6 & 20 31 48.58 & +41 27 39.99 & 6.3 & 2010 Jan 30, 21:19 & 28.8 \\ 
10944 & 2-6 & 20 31 26.47 & +41 31 19.39 & 12.9 & 2010 Feb 01, 10:57 & 28.7 \\ 
10945 & 3-6 & 20 32 04.37 & +41 34 58.80 & 12.9 & 2010 Feb 01, 19:13 & 28.2 \\ 
10946 & 4-6 & 20 32 42.26 & +41 38 38.21 & 12.9 & 2010 Feb 02, 09:40 & 28.7 \\ 
10947 & 5-6 & 20 33 20.15 & +41 42 17.61 & 12.9 & 2010 Feb 03, 22:14 & 27.6 \\ 
10948 & 6-6 & 20 33 58.05 & +41 45 57.02 & 12.9 & 2010 Feb 04, 20:40 & 27.6 \\ 
10949 & 5-5 & 20 33 39.63 & +41 35 10.69 & 12.9 & 2010 Feb 05, 04:41 & 27.8 \\ 
10950 & 4-5 & 20 33 01.73 & +41 31 31.29 & 12.9 & 2010 Feb 07, 17:04 & 27.8 \\ 
12099 & 1-2 & 20 32 06.48 & +40 59 12.30 & 6.7 & 2010 Feb 08, 17:13 & 6.0 \\ 
10951 & 3-5 & 20 32 23.84 & +41 27 51.88 & 19.6 & 2010 Feb 11, 13:54 & 29.6 \\ 
10952 & 2-4 & 20 32 05.42 & +41 17 05.55 & 20.2 & 2010 Feb 11, 22:27 & 29.6 \\ 
10953 & 2-3 & 20 32 24.90 & +41 09 58.63 & 20.6 & 2010 Feb 12, 07:05 & 29.3 \\ 
10954 & 5-3 & 20 34 18.58 & +41 20 56.85 & 20.4 & 2010 Feb 12, 15:26 & 29.5 \\ 
10955 & 4-3 & 20 33 40.68 & +41 17 17.44 & 27.4 & 2010 Feb 14, 18:37 & 27.8 \\ 
10956 & 3-3 & 20 33 02.79 & +41 13 38.04 & 27.4 & 2010 Feb 15, 02:41 & 28.1 \\ 
10957 & 4-4 & 20 33 21.21 & +41 24 24.36 & 27.4 & 2010 Feb 16, 15:50 & 29.4 \\ 
10958 & 3-4 & 20 32 43.32 & +41 20 44.96 & 27.4 & 2010 Feb 17, 00:40 & 29.4 \\ 
10959 & 5-4 & 20 33 59.10 & +41 28 03.77 & 27.4 & 2010 Feb 17, 09:15 & 28.5 \\ 
10960 & 4-2 & 20 34 00.16 & +41 10 10.52 & 27.4 & 2010 Feb 17, 21:55 & 28.6 \\ 
10961 & 3-2 & 20 33 22.27 & +41 06 31.12 & 27.4 & 2010 Feb 22, 06:55 & 30.2 \\ 
10962 & 2-5 & 20 31 45.95 & +41 24 12.47 & 27.4 & 2010 Feb 22, 15:45 & 29.8 \\ 
10963 & 5-2 & 20 34 38.05 & +41 13 49.93 & 27.4 & 2010 Feb 23, 00:13 & 29.8 \\ 
10964 & 2-2 & 20 32 44.37 & +41 02 51.71 & 27.4 & 2010 Feb 24, 14:36 & 30.1 \\ 
10965 & 6-5 & 20 34 17.52 & +41 38 50.10 & 27.4 & 2010 Feb 24, 23:34 & 30.0 \\ 
10966 & 6-4 & 20 34 37.00 & +41 31 43.18 & 36.7 & 2010 Mar 02, 04:41 & 29.6 \\ 
10967 & 6-3 & 20 34 56.47 & +41 24 36.26 & 36.7 & 2010 Mar 02, 13:15 & 29.2 \\ 
10968 & 6-2 & 20 35 15.95 & +41 17 29.34 & 36.7 & 2010 Mar 02, 21:35 & 29.2 \\ 
10969 & 1-1 & 20 32 25.95 & +40 52 05.38 & 39.2 & 2010 Mar 03, 05:56 & 29.1 \\ 
10970 & 2-1 & 20 33 03.85 & +40 55 44.79 & 41.7 & 2010 Mar 05, 16:04 & 29.9 \\ 
10971 & 3-1 & 20 33 41.74 & +40 59 24.20 & 45.2 & 2010 Mar 08, 21:25 & 30.6 \\ 
10972 & 4-1 & 20 34 19.63 & +41 03 03.60 & 45.7 & 2010 Mar 09, 20:12 & 28.0 \\ 
10973 & 5-1 & 20 34 57.53 & +41 06 43.01 & 46.0 & 2010 Mar 10, 04:40 & 27.8 \\ 
10974 & 6-1 & 20 35 35.42 & +41 10 22.42 & 46.2 & 2010 Mar 10, 12:35 & 27.8 \\ 
\tableline 
\end{tabular}
\newline
\footnotesize
{\bf Notes.} Observations are listed in date order. The aim points and roll angles are obtained from the satellite aspect solution before astrometric corrections are applied. Units of right ascension are hours, minutes, and seconds; units of declination are degrees, arcminutes, and arcseconds. ObsID~12099 is the second part of the observation of field 1-2, the first part of which, ObsID~10939, was interrupted to allow a spacecraft software reboot.
\end{center}
\end{table*}

%% file: columns.tex
\begin{table*}
\begin{center}
\tiny
\caption{List of columns in the {\it Chandra} Cygnus~OB2 Legacy Survey catalog.} 
\label{columns} 
\begin{tabular}{lcl}
\tableline 
Column Label & Units & Description \\
\tableline 
Number & ... & Source number used within the {\it Chandra} Cygnus OB2 Legacy Survey \\
Name & ... & IAU source name; prefix is CXOCOB2 J ({\it Chandra} X-ray Observatory Cygnus OB2) \\
RA$^a$ & deg & Right ascension (J2000) \\
Dec$^a$ & deg & Declination (J2000) \\
PosErr$^a$ & arcsec & 1$\sigma$ error circle around the source position (Right ascension, Declination) \\
PosType$^a$ & ... & Method used to estimate source position (see Section~\ref{s-positions}) \\
SuspectFlag & ... & Flag indicating that the source is coincident with a known PSF feature (see Section~\ref{s-trimming}) \\
\\
Significance$^b$ & ... & Source significance (calculated from the full band) \\
ProbNoSrc$^b$ & ... & Base 10 logarithm of the $p$-value$^c$ for no-source null hypothesis (minimum value, see Section~\ref{s-trimming})\\
Prob\_band$^b$ & ... & Band used for ProbNoSrc (minimum value, see Section~\ref{s-trimming})\\
Detect\_method$^c$ & ... & Method used for detecting the candidate source\\
\\
ExpTime\_Tot & s & Total exposure time in {\it all} observations\\
ExpTime\_Nom$^d$ & s & Total exposure time in all merged observations for photometry \\
ExpTime\_Frac$^d$ & ... & Fraction of total exposure time merged for photometry\\
NumObs\_Tot & ... & Total number of observations of the source \\
NumObs\_Nom$^d$ & ... & Total number of observations merged for photometry \\
\\
Theta\_Lo & arcmin & Smallest off-axis angle for merged observations (see also Fig.~\ref{offaxis_map})\\
Theta & arcmin & Average off-axis angle for merged observations \\
Theta\_Hi & arcmin & Largest off-axis angle for merged observations \\
PSF\_Frac & ... & Average PSF fraction (at 1.5 keV) for merged observations (see Section~\ref{s-apertures}) \\
SrcArea & pixel$^2$ & Average aperture area for merged observations (in pixels) \\
\\
ProbKS\_single & ... & Smallest $p$-value$^e$ for the non-variable null hypothesis from a single observation\\
ProbKS\_merge & ... & Smallest $p$-value$^e$ for the non-variable null hypothesis from all merged observations\\
&&  (see Section~\ref{s-variability}) \\
\\
SrcCnts\_full & counts & Observed counts in merged source apertures (full band) \\
SrcCnts\_soft & counts & Observed counts in merged source apertures (soft band) \\
SrcCnts\_hard & counts & Observed counts in merged source apertures (hard band) \\
\\
BkgCnts\_full & counts & Observed counts in merged background regions (full band) \\
BkgCnts\_soft & counts & Observed counts in merged background regions (soft band) \\
BkgCnts\_hard & counts & Observed counts in merged background regions (hard band) \\
BkgScaling & ... & Scaling of the background extraction region \\
\\
NetCnts\_full & counts & Net counts in merged source apertures (full band) \\
NetCnts\_soft & counts & Net counts in merged source apertures (soft band) \\
NetCnts\_hard & counts & Net counts in merged source apertures (hard band) \\
\\
MeanEffArea\_full & cm$^2$ count photon$^{-1}$ & Mean ARF$^f$ value (full band) \\
MeanEffArea\_soft & cm$^2$ count photon$^{-1}$ & Mean ARF$^f$ value (soft band) \\
MeanEffArea\_hard & cm$^2$ count photon$^{-1}$ & Mean ARF$^f$ value (hard band) \\
\\
MedianEnergy\_full & keV & Median energy$^g$ of the observed spectrum (full band) \\
MedianEnergy\_soft & keV & Median energy$^g$ of the observed spectrum (soft band) \\
MedianEnergy\_hard & keV & Median energy$^g$ of the observed spectrum (hard band) \\
\\
EnergyFlux\_full & erg cm$^{-2}$ s$^{-1}$ & Energy flux of the observed spectrum (full band) \\
EnergyFlux\_soft & erg cm$^{-2}$ s$^{-1}$ & Energy flux of the observed spectrum (soft band) \\
EnergyFlux\_hard & erg cm$^{-2}$ s$^{-1}$ & Energy flux of the observed spectrum (hard band) \\
\\
NetCnts\_full\_Lo & count & 1$\sigma$ lower bound on NetCounts\_full \\
NetCnts\_full\_Hi & count & 1$\sigma$ upper bound on NetCounts\_full \\
NetCnts\_soft\_Lo & count & 1$\sigma$ lower bound on NetCounts\_soft \\
NetCnts\_soft\_Hi & count & 1$\sigma$ upper bound on NetCounts\_soft \\
NetCnts\_hard\_Lo & count & 1$\sigma$ lower bound on NetCounts\_hard \\
NetCnts\_hard\_Hi & count & 1$\sigma$ upper bound on NetCounts\_hard \\
\tableline 
\end{tabular}
\end{center}
\footnotesize
{\bf Notes.} The catalog is sorted by right ascension. The suffixes ``\_full,'' ``\_soft," and ``\_hard" on names of photometric quantities designate the {\it full} (0.5--8 keV), {\it soft} (0.5--2 keV), and {\it hard} (2--8 keV) energy bands. \\
$^a$ Source position quantities (RA, Dec, PosErr, PosType) are computed using a subset of each source's extractions chosen to minimize the position uncertainty (see Section~\ref{s-positions}). \\
$^b$ Source significance quantities (Significance, ProbNoSrc, Prob\_band) are computed using a subset of each source's extractions chosen to maximize significance (see Section~\ref{s-trimming}). \\
$^c$ Detection methods: CW (CIAO {\sc wav detect}), PW ({\it PWdetect}), CPW (source detected by both CIAO {\sc wav detect} and {\it PWdetect}), EW ({\it Enhanced} {\sc wav detect}), and KS (previously known sources).\\
$^d$ Source photometric quantities are computed using a subset of each source's extractions (indicated by ExpTime\_Nom, ExpTime\_Frac, NumObs\_Nom) that balance the conflicting goals of minimizing photometric uncertainty and of avoiding photometric bias (see Section~\ref{s-fullextraction}). \\
$^e$ In statistical hypothesis testing, the $p$-value is the probability of obtaining a test statistic at least as extreme as the one that was actually observed when the null hypothesis is true. \\
$^f$ In {\it Chandra} ACIS data analysis the ARF incorporates both the effective area of the observatory and the fraction of the observation for which data were actually collected for the source. \\
$^g$ The median energy is the median energy of extracted events correct for the background \citep[see][Section 7.3]{broo10}.\\
\end{table*}